\documentclass[fleqn,twoside]{article}
\usepackage{espcrc2}

\usepackage{graphicx}
\usepackage[figuresright]{rotating}

\newcommand{\AmS}{{\protect\the\textfont2
  A\kern-.1667em\lower.5ex\hbox{M}\kern-.125emS}}

\newcommand{\be}{\begin{equation}}
\newcommand{\ee}{\end{equation}}
\newcommand{\bq}{\begin{eqnarray}}
\newcommand{\eq}{\end{eqnarray}}

\title{Non-Standard Time Reversal for Particle Multiplets\\ and
the Spin-Flavor Structure of Hadrons}

\author
{M.Anselmino\address[to]{Dip. di Fisica Teorica, Universit{\`a} 
di Torino 
and INFN, Sezione di Torino, 10125 Torino, Italy}, 
V. Barone\addressmark[to]\address{Di.S.T.A., Universit{\`a} 
del Piemonte Orientale 
``A.~Avogadro'', 15100 Alessandria, Italy}, 
A. Drago\address{Dip. di Fisica, Universit{\`a} di Ferrara 
and INFN, Sezione di Ferrara, 44100 Ferrara, Italy} and 
F. Murgia\address{Dip. di Fisica, Universit{\`a} 
di Cagliari
and INFN, Sezione di Cagliari, 09042 Monserrato (CA), 
Italy}
}

\begin{document}

\begin{abstract}
We show that a system of quarks interacting with chiral fields
provides a physical representation of a ``non-standard'' 
time reversal for particle multiplets proposed by Weinberg. 
As an application, we argue 
that, if the internal structure of hadrons 
is described by a chiral lagrangian, the 
so-called time-reversal-odd quark distribution functions
might not be forbidden 
by time-reversal invariance. 
\vspace{1pc}
\end{abstract}

\maketitle

The purpose of this note is to present a physical 
representation of a ``non-standard'' time reversal
for particle multiplets proposed 
by Weinberg in his  Quantum 
Field Theory book \cite{weinberg}. The physical system  we 
shall discuss 
is a chiral model such as 
the $\sigma$-model. An application to the spin-flavor 
structure of hadrons will be presented.  

Time-Reversal (TR) invariance is a fundamental constraint 
on many physical processes. Acting on a momentum and 
spin eigenstate $\vert \vec p, j_3 \rangle$, the 
TR operator $T$ gives
\be
T \vert \vec p,  j_3 \rangle = (-1)^{j - j_3} 
\vert - \vec p, - j_3 \rangle \,, 
\label{TR1}
\ee
where $j$ is the particle's spin, $j_3$ its third 
component, 
and an irrelevant phase has been omitted.   

Consider now a multiplet of particles labeled by 
some quantum number $a$. In the standard
realization of TR, the $T$ operator is  
taken to be diagonal in $a$: 
\be
T \vert \vec p,  j_3, a \rangle = (-1)^{j - j_3} 
\vert - \vec p, - j_3, a \rangle \,.  
\label{TR1b}
\ee
In \cite{weinberg} Weinberg explores 
a more general possibility, namely that $T$ may 
mix the multiplet components. Thus a non-diagonal 
finite matrix ${\mathcal T}_{ab}$ appears in (\ref{TR1b}), 
that becomes
\be
T \vert \vec p, j_3, a \rangle = 
(-1)^{j - j_3} 
\sum_b {\mathcal T}_{ab} \vert - \vec p, - j_3, b \rangle\,,  
\label{TR5}
\ee
Since $T$ is antiunitary, ${\mathcal T}$ must be unitary.
Weinberg proves that
the matrix ${\mathcal T}$ can be made block-diagonal
by a unitary transformation, 
with the blocks
being either simple phases, or 
at most $2 \times 2$ matrices of the form
\bq
\left(
\begin{array}{cc}
 0 & {\rm e}^{i\phi/2}  \\
 {\rm e}^{-i\phi/2} & 0  \\
\end{array}
\right) \,, 
\label{TR6}
\eq
where $\phi$ is a  real number.
In \cite{weinberg} no physical application of this ``unconventional''
TR is discussed. 
Our aim is to provide such an application.

Let us consider a SU(2) chiral lagrangian describing the  
interaction of a  fermionic field $\psi$ 
with two chiral fields $\sigma$ and $\vec \pi$ 
\bq
{\mathcal L} &=& i\bar \psi \gamma^{\mu}\partial_{\mu} \psi
       -g \bar \psi\left(\sigma
+i\gamma_5\vec\tau\cdot\vec\pi\right) \psi 
\nonumber \\
& &    +{1\over 2}{\left(\partial_\mu\sigma \right)}^2
       +{1\over 2}{\left(\partial_\mu\vec\pi\right)}^2
         -U\left(\sigma ,\vec\pi\right)   \, .
\label{TR7}
\eq

The field equations obtained from this lagrangian, 
in the mean field approximation, 
read \cite{cohen}:
\be
[i \gamma^\mu \partial_\mu -g(\sigma+i\gamma_5\vec\tau\cdot\vec\pi)]
\varphi(x)=0 \, ,
\label{TR8}
\ee
\be
-\nabla ^2\vec\pi + N_c i g \bar\varphi(x)\gamma_5\vec\tau\varphi(x)+
{\partial U\over \partial \vec\pi} =0 \, ,
\label{pion}
\ee
\be
-\nabla ^2\sigma + N_c  g \bar\varphi(x)\varphi(x)+
{\partial U\over \partial \sigma} =0 \, .
\label{sigma}
\ee

Under TR the Dirac equation becomes (the superscript ``t'' denotes transpose)
\be
[i \gamma^\mu \partial_\mu 
-g(\sigma-i\gamma_5\vec\tau\,^{\rm t} \cdot\vec\pi')]
\gamma_5 {\mathcal C} \varphi^*(\widetilde {x})=0 \, ,
\label{TR9}
\ee
where $\tilde x=(-t, \vec x)$, ${\mathcal C} = 
i \gamma_2 \gamma_0$ and we have denoted $\vec\pi'$ the time-reversed
pionic field. In absence    
of the pion field, the time-reversed solution would be, 
as usual, 
$\gamma_5 \, {\mathcal C} 
\varphi^*(\tilde x)$. But in (\ref{TR9}) the term containing $\vec \pi$
has changed and we need to specify how the pion field transforms
in order to define completely the time-reversed solution. Obviously
we have also to satisfy the 
time-reversed pion field equation (the equation
for the sigma field is invariant under TR), which reads:
\be
-\nabla ^2\vec\pi - 
N_c i g \bar\varphi(x)'\gamma_5\vec\tau\,^{\rm t}\varphi(x)'+
{\partial U\over \partial \vec\pi} =0 \, ,
\label{TRpion}
\ee
where we have denoted $\varphi(x)'$ the time-reversed quark field.

There are two ways to define a TR solution satisfying all 
equations. The first one amounts to
keeping the standard TR for the quarks and 
reversing the sign of the $x$ and $z$ components of the pionic
field under TR. This realization of TR
is generally used in systems in which pions are emitted or absorbed
by a fermion in a perturbative way. By inspecting eqs.(\ref{TR9}) and 
(\ref{TRpion}), we recognize the existence of another possible realization
of TR,
which consists in leaving the pionic field unchanged under TR and 
performing
an isospin rotation on quark field. Since
$\tau_2(-\vec\tau\,^{\rm t})\tau_2=\vec\tau$, the time-reversed
solution is ($ab$ are now isospin indices)
\be
T \varphi_a(k) = 
(\tau_2)_{ab} \gamma_5 \, {\mathcal C} \, \varphi_{b}^*(\tilde k)\,.  
\label{TR10}
\ee
The unitary isospin rotation $\tau_2$ is exactly 
of the form (\ref{TR6}) indicated by Weinberg, 
with $\phi=-\pi$. 

If we generalize the above lagrangian to SU(3), it 
is straightforward to show that there is no 
unitary $3 \times 3$ matrix that allows one to obtain 
the time-reversed solution. This agrees with 
Weinberg's conclusion that non-standard TR can mix 
at most two components of the particle multiplet. 

It is important to notice that the nucleon state built up from the 
chiral model that we are considering satisfies the usual TR properties. 
Starting from the mean-field solution, which has the hedgehog 
configuration $\vert h\rangle$,
we project 
out a state with definite spin and isospin \cite{many}
\begin{equation}
\vert J,J_3,  I_3\rangle \sim
\int d^3\Omega\,
{{\mathcal D}^{J\textstyle{*}}_{J_3,-I_3}}(\Omega )R(\Omega ) \vert h\rangle
\label{TR13}
\end{equation}
where ${\mathcal D}^J_{J_3,- I_3}(\Omega )$ is the 
familiar Wigner function and $R(\Omega )$ is the rotation operator.
The nucleon state built as in (\ref{TR13}) 
transforms in the standard way under TR,
i.e. according to eq.(\ref{TR1}), even though one exploits only
the invariance of $\vert h\rangle $ under the non-standard TR.  

In the literature some authors have introduced
the so-called ``TR-odd''
quark distribution functions \cite{sivers} to account 
for the single-spin asymmetries experimentally observed
in transversely polarised pion hadroproduction.
It is clear that if 
the TR operator acting on the quark 
field is the one given in (\ref{TR10})
the argument presented in \cite{collins} for the vanishing
of the TR-odd quark distribution functions can be circumvented and 
these quantities would not be forced to be zero due to TR even in
the absence of initial-state interactions.
The experimental measurement
of these distributions might clarify which realization of TR 
takes place inside nucleons.

\end{document}